\pdfoutput=1
\documentclass[journal]{IEEEtran}%
\usepackage{color}
\usepackage{amsmath}
\usepackage{multirow}
\usepackage{booktabs}
\usepackage{bm}
\usepackage{graphicx}
\usepackage{subfigure}
\usepackage{epstopdf}
\usepackage{cite}
\usepackage{amssymb}
\usepackage{threeparttable}
\usepackage{algorithm}
\usepackage{algorithmic}
\usepackage{enumerate}

\usepackage{url}

\ifCLASSINFOpdf

\else

\fi

\begin{document}

\title{Maximum Production Point Tracking of a High-Temperature Power-to-Gas System: \\A Dynamic-Model-Based Study}

\author{Xuetao~Xing,~\IEEEmembership{Student~Member,~IEEE,~}%Student~Member,~IEEE,
        Jin~Lin,~\IEEEmembership{Member,~IEEE,}
        Yonghua~Song,~\IEEEmembership{Fellow,~IEEE,}
        Qiang~Hu% <-this % stops a space
%\thanks{This work was partly supported by National High-Technology Research and Development Program (863 Program) of China (2015AA050102).}
%, International S\&T Cooperation Program of China (2014DFG62670), National Natural Science Foundation of China (51207077, 51261130472, 51577096), and China Postdoctoral Science Foundation (2015M580097).}
\thanks{Xuetao Xing and Jin Lin are with the State Key Laboratory of Control and Simulation of Power Systems and Generation Equipment, Department of Electrical Engineering, Tsinghua University, Beijing 100084, China (e-mail: linjin@tsinghua.edu.cn).}
\thanks{Yonghua Song is with the Department of Electrical and Computer Engineering, University of Macau, Macau, China, and with the Department of Electrical Engineering, Tsinghua University, Beijing 100084, China.}
\thanks{Qiang Hu is with Tsinghua-Sichuan Energy Internet Research Institute, Chengdu 610213, China.}}% <-this % stops a space

%\markboth{XXX Journal of \LaTeX\ Class Files,~Vol.~14, No.~8, August~2015}%

\maketitle
\begin{abstract}
  Power-to-gas (P2G) can be employed to balance renewable generation because of its feasibility to operate at fluctuating loading power.
  The fluctuating operation of low-temperature P2G loads can be achieved by controlling the electrolysis current alone.
  However, this method does not apply to high-temperature P2G (HT-P2G) technology with auxiliary parameters such as temperature and feed rates: Such parameters need simultaneous coordination with current due to their great impact on conversion efficiency.
  To improve the system performance of HT-P2G while tracking the dynamic power input, this paper proposes a maximum production point tracking (MPPT) strategy and coordinates the current, temperature and feed rates together.
  In addition, a comprehensive dynamic model of an HT-P2G plant is established to test the performance of the proposed MPPT strategy, which is absent in previous studies that focused on steady states.
  The case study suggests that the MPPT operation responds to the external load command rapidly even though the internal transition and stabilization cost a few minutes.
  Moreover, the conversion efficiency and available loading capacity are both improved, which is definitely beneficial in the long run.
\end{abstract}

\IEEEpeerreviewmaketitle

\section*{Nomenclature}
%The main notations of this paper are listed below; other symbols are defined as required.
\subsection{Constants and Variables}
\addcontentsline{toc}{section}{Nomenclature}
\begin{IEEEdescription}[\IEEEusemathlabelsep\IEEEsetlabelwidth{$V_1,V_2$}]
\item[$F$] Faraday constant.
\item[$M$] Molar mass.
\item[$T,p$] Temperature, pressure.
%\item[$M$] Molar mass.
%\item[$h$] Specific enthalpy.
\item[$c,C$] Specific heat capacity, heat capacity (isobaric).
\item[$U$] Voltage or overvoltage.
\item[$I,i$] Current, current density.
\item[$w$] Mass flow rate of the gas mixture.
\item[$\pi$] Feed factor.
%\item[$\omega$] Mass fraction.
\item[$P,Q$] Active power, reactive power.
\item[$\eta$] Efficiency.
\item[$\tau$] Time constant.
\item[$N$] Number.
\item[$R,\mathbb{C},\Phi$] Resistance, capacitance, flux linkage.
\item[$J,\Omega$] Angular mass, angular velocity.
\item[$s$] Laplace variable.
\item[$k,K$] Dimensionless coefficient, PID parameter.
\item[$\bm{p},\bm{w},\bm{\pi}$] $\begin{bmatrix} p_{\rm H_2O}\! & p_{\rm H_2}\! & p_{\rm O_2} \end{bmatrix} ^{\rm T}$, $\begin{bmatrix} w_{\rm ca} & w_{\rm an} \end{bmatrix} ^{\rm T}$, $\begin{bmatrix} \pi_{\rm ca} & \pi_{\rm an} \end{bmatrix} ^{\rm T}$.
\end{IEEEdescription}

\subsection{Subscripts}
\addcontentsline{toc}{section}{Nomenclature}
\begin{IEEEdescription}[\IEEEusemathlabelsep\IEEEsetlabelwidth{$V_1,V_2,V_3,V_4$}]
\item[$\rm ca,an,el$] Cathode, anode, electrolysis.
\item[$\rm rev,th$] Reversible, thermo-neutral (voltage).
\item[$\rm act,con,ohm$] Activation, concentration, ohmic (overvoltage).
\item[$\rm war,rea,rec$] Warming, reaction, heat recovery.
\item[$\rm pre,fur,cel$] Preheater, furnace, (SOC) cell.
\item[$\rm pum,com,cvt$] Pump, compressor, converter.
\item[$\rm flt,nom,gen$] Filter, nominal, generate.
\item[$\rm mot,fra$] Motor, fraction.
\item[$\rm vap,liq,amb$] Vaporization, liquid, ambient.
\item[$\rm thr,opt,ref$] Threshold, optimal, reference.
\item[$\rm in,out$] Inlet, outlet position of the furnace.
\item[$\rm p,i,d$] Proportion, integration, differentiation.
\end{IEEEdescription}

\section{Introduction}

\IEEEPARstart{T}{he} penetration and utilization ratios of renewable generation sources are limited by the adverse grid-side impacts of their naturally intermittent outputs.
In this context, the emerging technology of power-to-gas (P2G) provides the necessary flexibility to complement source-side uncontrollability and thus facilitate renewable integration \cite{Clegg_P2G_2015}.

The idea is that P2G can operate at various loading conditions and that the coupling gas demands (injected into storage tanks or directly into gas pipelines) are typically flexible \cite{Petipas_HTE_VariousLoads_2013, Kopp_EnergiparkMainz_2017}.
Therefore, the loading power of P2G can rapidly track the fluctuating dispatch commands, which can be generated by the grid regulator to moderate the peak power and balance the uncontrollable renewable energy \cite{Frank_rSOC_op_2018,Khani_SchedulingP2G_2017}.
In this manner, the renewable energy that may otherwise be curtailed can be used to produce easily applicable fuel gases.
%Specifically, the P2G system can be co-located with a renewable power plant like an on-site energy storage unit to buffer and correct the output profile \cite{Kopp_EnergiparkMainz_2017}, or it can be deployed as a flexible load to provide power balancing in a microgrid or an active distribution network \cite{Khani_SchedulingP2G_2017}.
Several pilot projects of P2G participating renewable energy regulation have been realized worldwide \cite{Kopp_EnergiparkMainz_2017, Gahleitner_P2Gplants_2013}.

For commercialized low-temperature P2G technologies (typically below $100^\circ{\rm C}$ \cite{Gahleitner_P2Gplants_2013}), the system loading power is almost totally attributed to the electric energy consumed by the electrolysis process, where $\rm H_2O$ splits into $\rm H_2$ and $\rm O_2$.
%With optional methanation reactor and $\rm CO_2$ injection, $\rm H_2$ can be further converted into $\rm CH_4$ \cite{Lehner_P2G_2014}.
By altering the electrolysis current with a necessary power converter, the loading power can respond to the command signal rapidly, which has been verified to be fast enough to participate in automatic generation control (AGC) \cite{Archambault_AGC_2015}.

\begin{figure}[!t]
	\centering
	\includegraphics[width=0.49\textwidth]{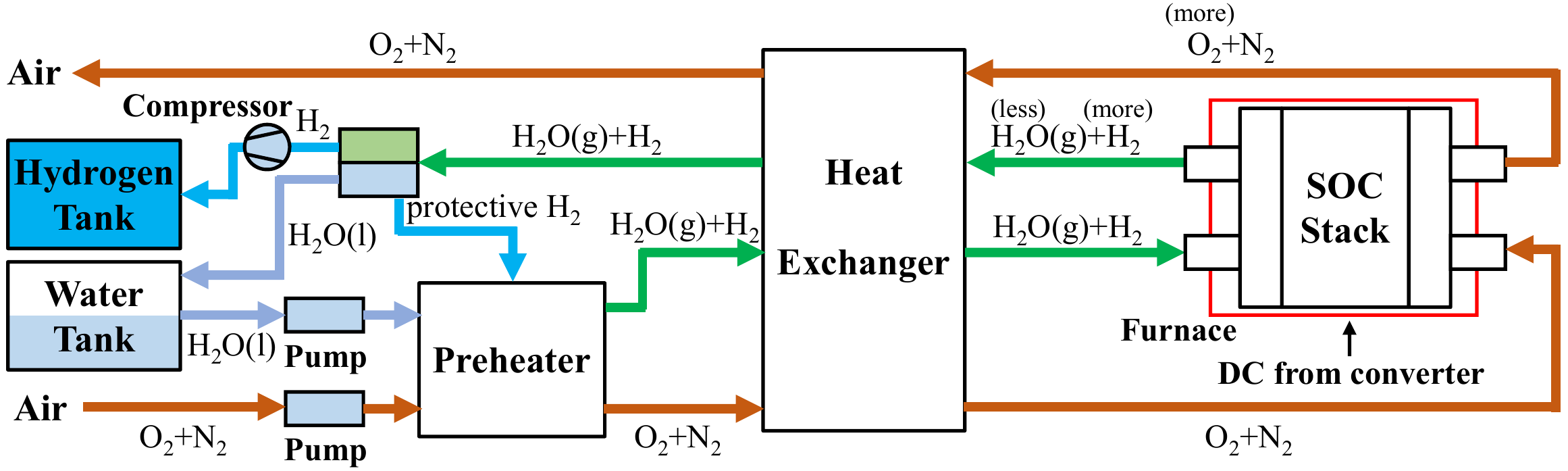}\\
	\caption{General structure of an HT-P2G system \cite{Frank_rSOC_op_2018}, \cite{Pan_SOECsystem_experiment_2017}.}\label{fig:P2G_plant}
\end{figure}

High-temperature P2G (HT-P2G) is a developing technology that employs solid oxide cells (SOCs) to electrolyze vaporous water, as shown in Fig. \ref{fig:P2G_plant}.
Auxiliary modules such as the preheater, the heat exchanger and the furnace are typically required for necessary heating.
Relative to traditional low-temperature P2G technologies, HT-P2G can achieve higher conversion efficiency because the electrolysis reaction is facilitated both thermodynamically and kinetically at elevated temperatures (typically above $750^\circ{\rm C}$) \cite{Udagawa_SOECmodel_2007}.
Nevertheless, the traditional current-controlled-alone method of low-temperature P2G is not well suited to HT-P2G:
Despite the fact that the controlling current is sufficient for tracking the desired loading power \cite{Petipas_HTE_VariousLoads_2013,Frank_rSOC_op_2018,Udagawa_SOECmodel_2007}, the system performance of HT-P2G is far from optimal without the simultaneous coordination of other HT-P2G parameters such as stack temperature and feed flow rates \cite{Petipas_HTE_VariousLoads_2013}, which affects the power consumption of auxiliary modules and thus the conversion efficiency.
This topic is the focus of this paper.

The fundamentals of HT-P2G operation, such as models of SOCs and auxiliary modules, have long been described in the literature.
Reference \cite{Udagawa_SOECmodel_2007} presented a one-dimensional SOC model to evaluate the cell voltage from the electrolysis current with a reversible voltage and various overvoltages, which was validated experimentally in \cite{Kazempoor_SOECmodel_2014}.
Reference \cite{Padulles_SOFC_TransferFunction_2000} built a lumped model with transfer functions to succinctly describe the electrical dynamics of an SOC stack.
As for models of necessary auxiliaries including the power converter, furnace (or heater), pump and compressor, such information can also be found in the literature \cite{Gasser_PEMFC_SystemModel_2006, Petipas_HTE_VariousLoads_2013, Milewski_AuxiliaryDynamic_2014}.

Based on the above fundamental studies, research papers have discussed the parameter optimization of a running HT-P2G system.
Reference \cite{Brien_NuclearHTE_2010} analyzed the correlation between the overall $\rm H_2$ production efficiency and the operating parameters, such as the temperature, production rate, steam utilization and air sweep, based on a simulated HT-P2G system coupled to a nuclear reactor.
Reference \cite{Frank_rSOC_op_2018} focused on the impacts of the $\rm H_2$ recirculation rate and steam utilization on the system efficiency of a reversible SOC plant.
Reference \cite{Wang_P2M_opt_2018} calculated the optimal HT-P2G operating points at various loading conditions using a multi-objective optimizer and conducted further analysis with the obtained scatter diagrams.
Reference \cite{Luo_Excergy_SOECsystem_2018} performed a simulation-based investigation of the energy efficiency and its dependence on feedstock composition, cathode feed rate, temperature, SOC pressure, among other parameters.
As a result of these HT-P2G optimization studies, some principles to achieve higher system performance have been commonly accepted, such as reducing the air sweep and enhancing the steam utilization.
Nevertheless, these studies mostly focused only on the improvement of steady-state performance, while the investigation of dynamic models has not been presented.
In other words, the research to describe the plant's transient behaviors considering inherent inertia is currently insufficient, which is essentially required in developing advanced control strategies to track fluctuating load commands. 

In this paper, the optimization of operating parameters is combined with a comprehensive dynamic model of an HT-P2G system that describes its transient behaviors.
Specifically, a maximum production point tracking (MPPT) control strategy is proposed to enable the HT-P2G plant to rapidly track the desired loading power profile and simultaneously maximize the steady-state energy conversion efficiency by coordinating various auxiliary modules properly.
The effects of the proposed strategy are validated with a practical case where an HT-P2G plant is dispatched to implement AGC as a flexible load.

This paper primarily achieves the following contributions:
1) An integrated HT-P2G dynamic model is presented to describe a system's dependence on temperature and feed flow parameters considering auxiliary but necessary energy consumptions and to depict its transient behaviors with multi-scale dynamics of several coupled domains (such as the electrical domain, thermodynamics and hydromechanics);
2) An MPPT control strategy is proposed to coordinate the temperature and feed flow parameters other than electrolysis current, which shows a fast response speed to track the loading power command in the short term and significantly beneficial effects by increasing the production efficiency and available capacity in the long run.

%The remainder of the paper is organized as follows.
%Section II describes the mathematical model of multi-terminal LPC in detail, where the IDD method is introduced.
%LPC-ADN is comprehensively modeled in Section III.
%Section IV develops an MPC strategy based on rolling horizon optimization.
%The case studies are conducted in Section V to verify the proposed model and framework.
%Finally, the conclusions are given in Section VI.

\section{Dynamic Description of an HT-P2G System}
%Starting with necessary fundamentals such as typical structures and chemical thermodynamics, this section presents a general analysis of an HTE based PtG plant in terms of energy balance and energy conversion.
\subsection{Overall Model of the System}
\begin{figure}[!t]
  \centering
  \includegraphics[width=0.49\textwidth]{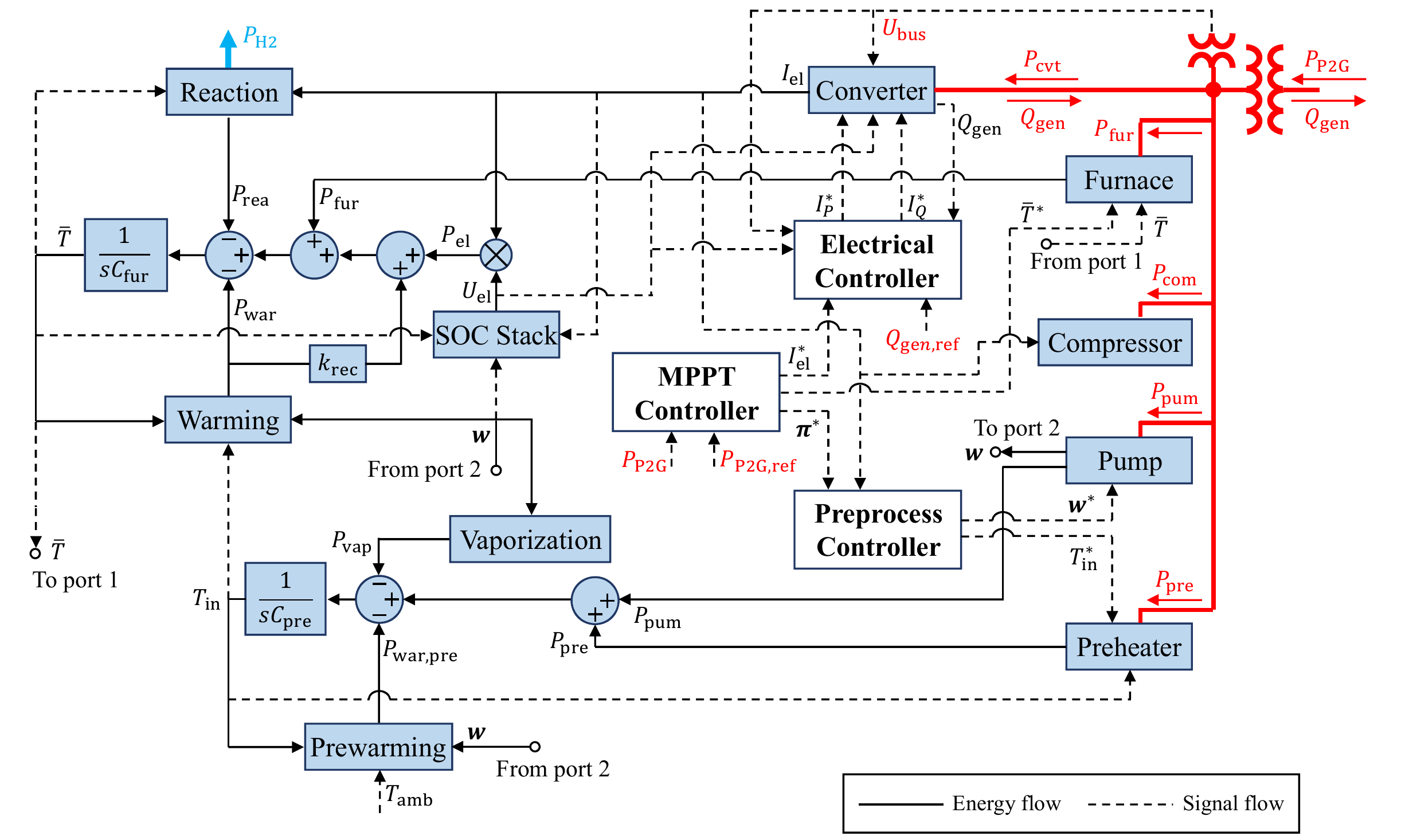}\\
  \caption{Overall dynamic model of an HT-P2G system (the involved submodels are elaborated in Fig. \ref{fig:EnergySinkDiagram}, Fig. \ref{fig:SOCstackDiagram} and Fig. \ref{fig:AuxiliaryModuleDiagram}; the controller designs are presented in Fig. \ref{fig:MPPTControllerDiagram} and Fig. \ref{fig:ControllerDiagram}).}\label{fig:SystemDiagram}
\end{figure}
As shown in Fig. \ref{fig:P2G_plant}, a typical HT-P2G system consists of an SOC stack where water electrolysis occurs in addition to necessary auxiliary modules such as converter, preheater, pump group, furnace and compressor \cite{Zhang_HTEsystem_experiment_2015,Petipas_HTE_VariousLoads_2013,Wang_P2M_opt_2018,Frank_rSOC_op_2018}.
Driven by the pump group, $\rm H_2O$ (mass flow rate $w_{\rm ca}$, with some protective $\rm H_2$ \cite{Udagawa_SOECmodel_2007,Frank_rSOC_op_2018}) and sweep air (mass flow rate $w_{\rm an}$) first goes through the preheater that vaporizes water and then enters the furnace, where steam and air are further heated and fed into the cathode channels and anode channels of the SOC stack, respectively.
Inside each SOC, steam is converted into $\rm H_2$ at average SOC temperature $\bar{T}$ with the presence of DC electrolysis current $I_{\rm el}$ provided by the converter.
After leaving the furnace, the $\rm H_2$-rich humid gas is condensed for purification and then pressurized to $p_{\rm com}$ by the compressor as the final product.

The whole system model of Fig. \ref{fig:P2G_plant} is shown in Fig. \ref{fig:SystemDiagram}.
The SOC productive condition $(I_{\rm el},\bar{T},\bm{w})$ is ensured by the controlled auxiliary modules, whose energy consumptions add up to the system loading power $P_{\rm P2G}$.
In addition, the energy balances regarding the heating processes within furnace and preheater are modeled to evaluate the variations of $\bar{T}$ and $T_{\rm in}$, respectively.
Note that the solid blocks in Fig. \ref{fig:SystemDiagram} represent inherent elements or physical relations, while the hollow blocks represent artificially designed controllers.
We provide a closer view of this model in the following subsections.

\subsection{Dynamic Models of the Energy Balances}\label{subsec:model_energyBalance}
According to \cite{Petipas_HTE_VariousLoads_2013}, the furnace volume energy balance can be described as
\begin{equation}\label{eq:dT_bar}
  C_{\rm fur} \frac{{\rm d}\bar{T}}{{\rm d}t} \doteq U_{\rm el}I_{\rm el} + P_{\rm fur} + k_{\rm rec} P_{\rm war} - P_{\rm war} - P_{\rm rea}.
\end{equation}
The energy used for feedstock warming ($P_{\rm war}$) and the electrolysis reaction ($P_{\rm rea}$) is subtracted from the energy provided by the furnace ($P_{\rm fur}$), DC current ($U_{\rm el}I_{\rm el}$) and heat recovery ($k_{\rm rec} P_{\rm war}$) to obtain the net input energy, which accounts for the change in $\bar{T}$.
Here, $k_{\rm rec}$ is the heat recovery coefficient, representing the fraction of exhaust heat recovered for inlet gas warming.
In the preprocess, the energy is primarily distributed to warming ($P_{\rm war,pre}$) and vaporization ($P_{\rm vap}$), and the energy balance can be analogously formulated as
\begin{equation}\label{eq:dT_in}
  C_{\rm pre} \frac{{\rm d}T_{\rm in}}{{\rm d}t} \doteq P_{\rm pre} + P_{\rm pum} - P_{\rm war,pre} - P_{\rm vap}.
\end{equation}
Note that $C_{\rm fur}$ and $C_{\rm pre}$ are equivalent heat capacities that can be measured experimentally.

\begin{figure}[!t]
  \centering
  \includegraphics[width=0.49\textwidth]{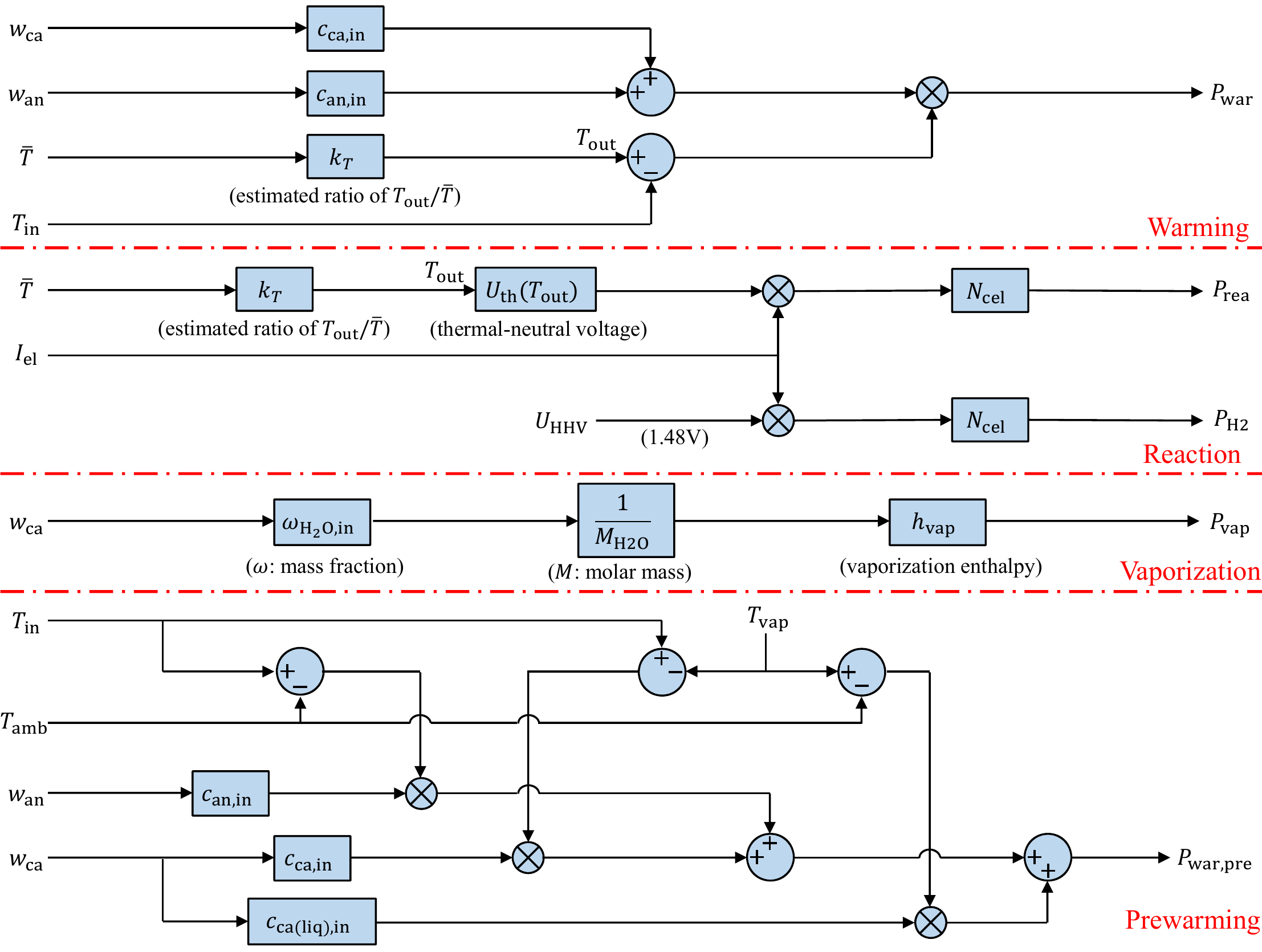}\\
  \caption{Dynamic models of primary energy sinks (based on \cite{Petipas_HTE_VariousLoads_2013}).}\label{fig:EnergySinkDiagram}
\end{figure}

Relations \eqref{eq:dT_bar} and \eqref{eq:dT_in} are included in Fig. \ref{fig:SystemDiagram}, and detailed models of the aforementioned energy sinks are depicted in Fig. \ref{fig:EnergySinkDiagram}.
Note that $P_{\rm war}$ and $P_{\rm war,pre}$ are calculated from the corresponding heat capacities and temperature increments, while $P_{\rm rea}$ and $P_{\rm vap}$ are calculated from the enthalpy changes in the electrolysis or vaporization process.

\subsection{Dynamic Model of the SOC Stack}\label{subsec:model_SOC}
\begin{figure}[!t]
  \centering
  \includegraphics[width=0.49\textwidth]{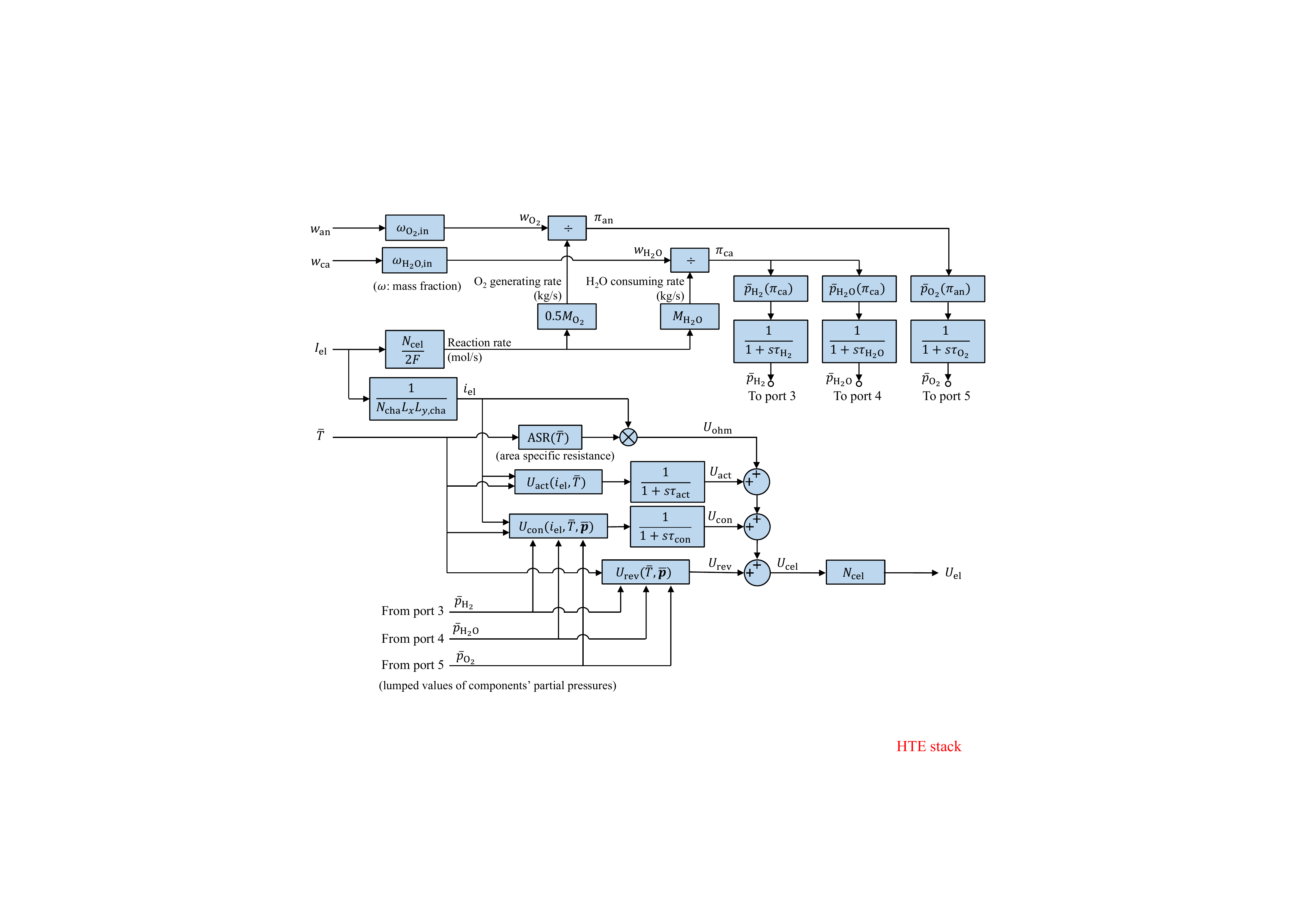}\\
  \caption{Dynamic model of the SOC stack (based on \cite{Padulles_SOFC_TransferFunction_2000}, \cite{Correa_overvoltageDynamic_2004} and \cite{Xing_HTE_2017}).}\label{fig:SOCstackDiagram}
\end{figure}

According to \cite{Padulles_SOFC_TransferFunction_2000}, the SOC stack can be modeled as shown in Fig. \ref{fig:SOCstackDiagram}.
This model outputs the dynamic behavior of stack voltage $U_{\rm el}$ based on $I_{\rm el}$, $\bar{T}$ and the feedstock flow rates $w_{\rm ca}$ and $w_{\rm an}$.
The cell voltage $U_{\rm cel}(i_{\rm el},\bar{T},\bar{\bm{p}})$ is calculated by totaling the reversible voltage ($U_{\rm rev}$) and various overvoltages ($U_{\rm ohm}$, $U_{\rm act}$ and $U_{\rm con}$).
The average partial pressures $p_{\rm H_2O}$, $p_{\rm H_2}$ and $p_{\rm O_2}$ are determined by the feed factors $\pi_{\rm ca}$ and $\pi_{\rm an}$, which represent the ratios of actual provided $\rm H_2O$ and $\rm O_2$ to their reaction-correlated quantities in the channels of cathode and anode, respectively.
(Alternatively, some papers use the steam conversion rate and air ratio, which are actually $\frac{1}{\pi_{\rm ca}}$ and $\pi_{\rm an}$, respectively \cite{Petipas_HTE_VariousLoads_2013}.)
Note that the first-order delays of $\tau_{\rm act}$ and $\tau_{\rm con}$ arise from the cell's electrochemical double-layer phenomenon \cite{Correa_overvoltageDynamic_2004} and that the pressure inertia described by $\tau_{\rm H_2O}$, $\tau_{\rm H_2}$ and $\tau_{\rm O_2}$ arises from the finite gas flow rates \cite{Padulles_SOFC_TransferFunction_2000,Xing_HTE_2017}.
Note that the model of Fig. \ref{fig:SOCstackDiagram} is reversible: its output can still describe the SOC stack voltage when $I_{\rm el}<0$ (fuel cell mode); only the overvoltages $U_{\rm ohm}$, $U_{\rm act}$ and $U_{\rm con}$ are negative values.

\subsection{Dynamic Models of the Auxiliary Modules}\label{subsec:model_auxiliary}
\begin{figure}[!t]
  \centering
  \includegraphics[width=0.49\textwidth]{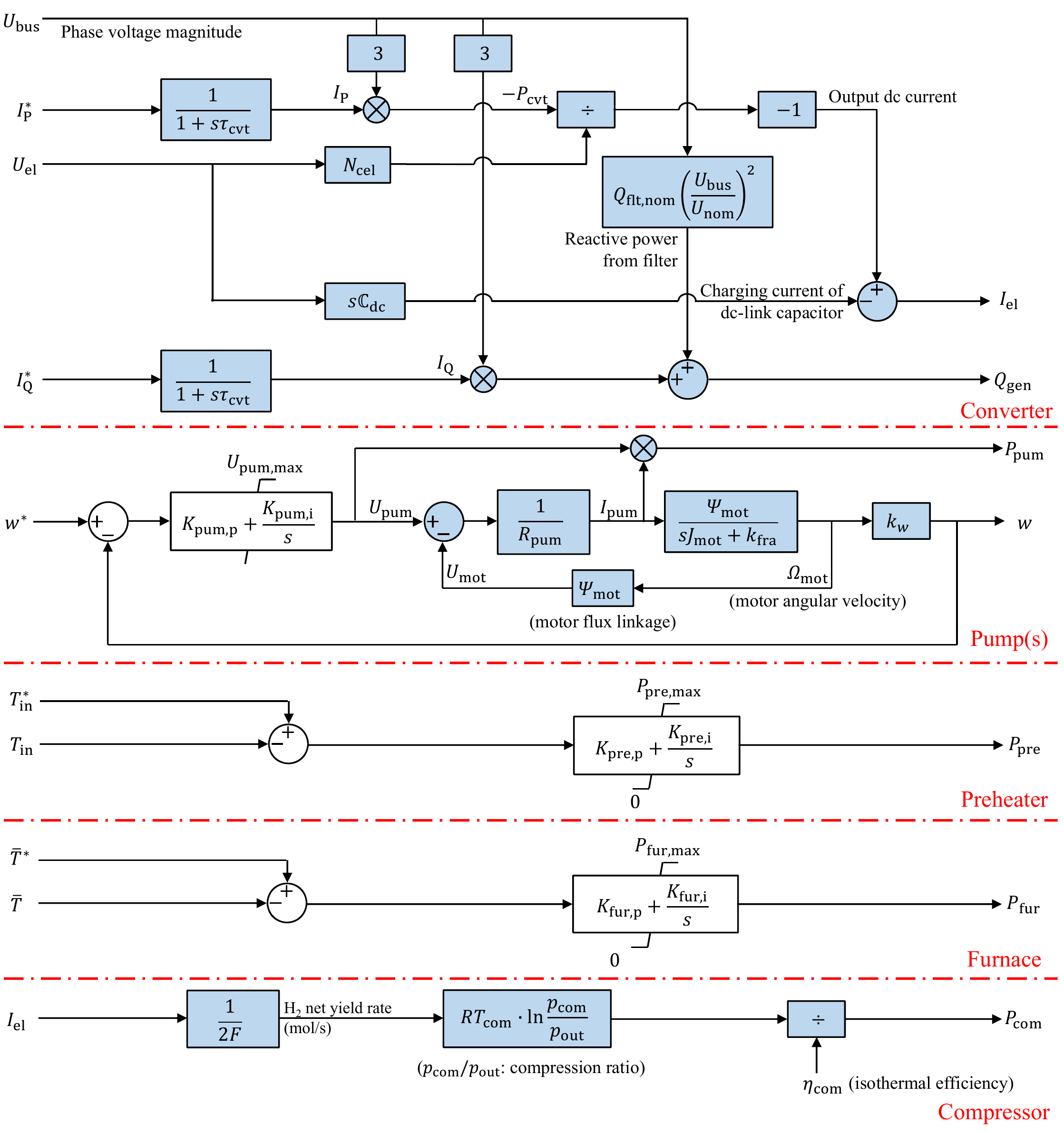}\\
  \caption{Dynamic models of various auxiliary modules (based on \cite{Kirubakaran_FC_converter_2009}, \cite{Gasser_PEMFC_SystemModel_2006}, \cite{Yu_furnaceControl_2011} and \cite{Meier_PEMandSOEC_2014}).}\label{fig:AuxiliaryModuleDiagram}
\end{figure}

The primary auxiliary modules of HT-P2G can be modeled as Fig. \ref{fig:AuxiliaryModuleDiagram}.
Below is a detailed explanation.

\subsubsection{Converter}
According to \cite{Kirubakaran_FC_converter_2009}, a three-phase PWM converter is preferred for medium- and high-power SOC applications, owing to its simple control and bidirectional power flow.
Here, we simply employ a first-order delay of $\tau_{\rm cvt}$ to model the rapid control of active and reactive components of the grid-side current, $I_{\rm P}$ and $I_{\rm Q}$, which can be realized with classic decoupled $dq$-current control loops and the phase-locked loop of grid voltage \cite{Zhao_EnergyStorageControl_2018,Clark_GE_WindTurbine_2010}.
Then, the converter's reactive output $Q_{\rm gen}$, including the reactive generation of the filter, can be derived from $I_{\rm Q}$.
On the DC side, the electrolysis current $I_{\rm el}$ can be obtained by extracting the charging current of the DC link capacitor $\mathbb{C}_{\rm dc}$ from the total DC current derived from $I_{\rm P}$.
\subsubsection{Pump}
The DC-motor-based model in \cite{Gasser_PEMFC_SystemModel_2006} is employed to describe the metering pumps that drive the feedstock flows in both the cathode and anode channels.
As shown in Fig. \ref{fig:AuxiliaryModuleDiagram}, the flow rate $w_{\rm ca}$ (or $w_{\rm an}$) is controlled by a built-in fast current controller.
\subsubsection{Preheater and Furnace}
Employing the convectional design of heaters \cite{Yu_furnaceControl_2011}, the preheater and the furnace are regarded as electric heaters with built-in PI controllers to track the desired $T_{\rm in}$ and $\bar{T}$, respectively, as shown in Fig. \ref{fig:AuxiliaryModuleDiagram}.
\subsubsection{Compressor}
According to \cite{Meier_PEMandSOEC_2014}, the compressor's energy consumption $P_{\rm com}$ can be modeled by multiplying the molar flow of the output $\rm H_2$ and the logarithm of the compression ratio ${\rm ln}\frac{p_{\rm com}}{p_{\rm out}}$, considering its isothermal efficiency $\eta_{\rm com}$.

\section{Maximum Production Point Tracking}
%Based on the energy analysis above, this section studies the model as well as the optimization of steady-state operating point of an HTE based PtG plant, which plays an important role in the design of its local controller.

\subsection{Maximum Production Point (MPP) Solver}\label{subsec:MPP_solver}
A comprehensive steady-state model of the HT-P2G system is required to find its maximum production point (MPP) where the conversion efficiency $\eta_{\rm P2G}$ is optimized.
First, the total energy consumption should meet the desired loading power $P_{\rm P2G,ref}$:
\begin{equation}\label{eq:P_P2G_ref}
  \underline{U_{\rm el}}I_{\rm el} + P_{\rm fur} + P_{\rm pre} + P_{\rm pum} + \underline{P_{\rm com}} = P_{\rm P2G,ref}
\end{equation}
Furthermore, the steady-state equations of \eqref{eq:dT_bar} and \eqref{eq:dT_in} should be fulfilled:
\begin{equation}\label{eq:steady_T_bar}
  \underline{U_{\rm el}}I_{\rm el} + P_{\rm fur} + k_{\rm rec} \underline{P_{\rm war}} - \underline{P_{\rm war}} - \underline{P_{\rm rea}} = 0
\end{equation}
\begin{equation}\label{eq:steady_T_in}
  P_{\rm pre} + P_{\rm pum} - \underline{P_{\rm war,pre}} - \underline{P_{\rm vap}} = 0
\end{equation}
In addition, for steady operating states, there exists a maximum electrolysis current $I'_{\rm el,max}$ at given $P_{\rm fur}$ and $\bm{\pi}$ to ensure that the feedstock is not excessive and thus can be heated up to at least the lowest electrolysis temperature $T_{\rm min}$ at the stack entrance.
In other words, $I'_{\rm el,max}$, which represents the maximum feeding rate affordable by the current furnace power, can be obtained from the measured steady-state $I_{\rm el}$ when the stack entrance temperature just reaches $T_{\rm min}$.
\begin{equation}\label{eq:Ip_el_max}
  I_{\rm el} \leq I'_{\rm el,max}(P_{\rm fur},\bm{\pi})
\end{equation}
Additionally, the primary parameters have upper and lower limits to ensure the safe operation of the SOC stack:
\begin{equation}\label{eq:T_bar}
  T_{\rm min} \leq \bar{T} \leq T_{\rm max}
\end{equation}
\begin{equation}\label{eq:I_el}
  I_{\rm el,min} \leq I_{\rm el} \leq I_{\rm el,max}
\end{equation}
\begin{equation}\label{eq:pis}
  \pi_{\rm ca} > 1, \pi_{\rm an} \geq 0
\end{equation}
Then, the MPP can be acquired by maximizing $\eta_{\rm P2G}$:
\begin{equation}\label{eq:efficiency}
  \max_{I_{\rm el},\bar{T},\bm{\pi}} \frac{N_{\rm cel} I_{\rm el} U_{\rm HHV}}{P_{\rm P2G,ref}}
\end{equation}
within the constraints of \eqref{eq:P_P2G_ref}-\eqref{eq:pis}.
Note that the aforementioned underlined variables $\underline{U_{\rm el}}$, $\underline{P_{\rm con}}$, $\underline{P_{\rm com}}$, $\underline{P_{\rm war}}$, $\underline{P_{\rm war,pre}}$, $\underline{P_{\rm rea}}$ and $\underline{P_{\rm vap}}$ can all be simply derived from the operating parameters $(I_{\rm el},\bar{T},\bm{\pi})$ based on the steady states of the dynamic models described in Section \ref{subsec:model_energyBalance}, \ref{subsec:model_SOC} and \ref{subsec:model_auxiliary}, as elaborated in Appendix \ref{appsec:Underline}.
By employing the interior-point algorithm, \eqref{eq:P_P2G_ref}-\eqref{eq:efficiency} are sufficient for solving MPPs numerically at various loading conditions $P_{\rm P2G,ref}$.

\subsection{Visualized Explanations of MPP Locating}
Based on the numerical MPP solutions obtained in Section \ref{subsec:MPP_solver}, patterns can be found to help us understand the MPP problem in a visual manner:

1) $\pi_{\rm an}$ is always zero (i.e., no sweep air) at MPPs.
A larger $\pi_{\rm an}$ (i.e., a large sweep air flow rate) increases the required energy for heating but also benefits the system by lowering $\bar{p}_{\rm O_2}$, $U_{\rm el}$ and thus the required energy for the converter.
However, the drop in $U_{\rm el}$ due to $\pi_{\rm an}$ is essentially negligible because the oxygen fraction is included as a logarithm when calculating $U_{\rm el}$;  it ranges $(0.21,1]$ but never approaches zero when $\pi_{\rm an}$ rises.
In this context, the sweep air does not detectably benefit the system performance, which is consistent with the analyses in \cite{Frank_rSOC_op_2018} and \cite{Brien_NuclearHTE_2010}.

2) $I_{\rm el}$ is always kept at $I'_{\rm el,max}$ at MPPs.
In other words, the constraint \eqref{eq:Ip_el_max} is always active at optimal solutions, and the furnace power $P_{\rm fur}$ is used at its fullest.
If not, one can always lower the parameter $\bar{T}$ or $\pi_{\rm ca}$ (but maintain $I_{\rm el}$) appropriately to reduce actual $P_{\rm fur}$ and then fulfill the equality sign of \eqref{eq:Ip_el_max}; the new steady state has a lowered $P_{\rm P2G}$ and thus a higher $\eta_{\rm P2G}$.
We designate \eqref{eq:Ip_el_max} as the ``furnace constraint'' (FC) for convenience.

\begin{figure}[!t]
  \centering
  \includegraphics[width=0.49\textwidth]{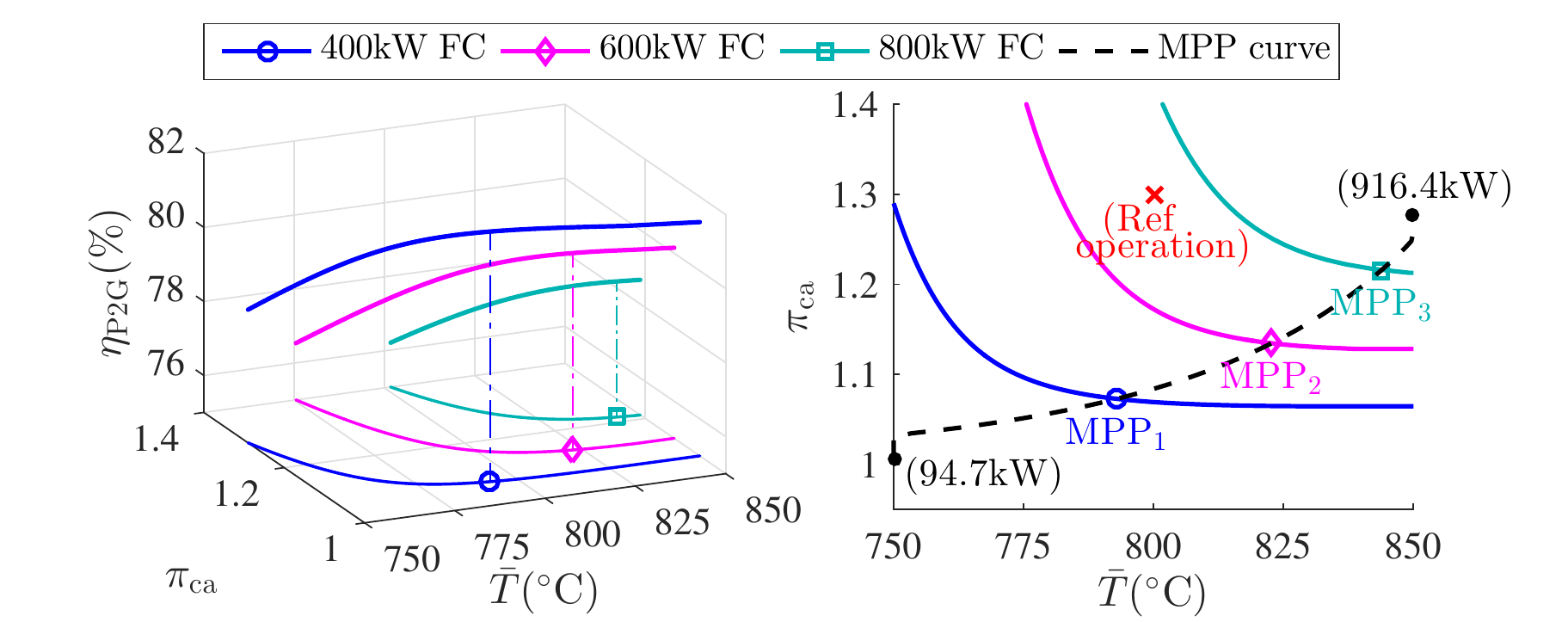}\\
  \caption{Solving the optimal operating states at various $P_{\rm P2G,ref}$ to obtain the MPP curve.}\label{fig:MPP_curve}
\end{figure}
Based on the two patterns above, we can understand the MPP problem \eqref{eq:P_P2G_ref}-\eqref{eq:efficiency} as ``finding the optimal $\bar{T}$ and $\pi_{\rm ca}$ that maximize $I_{\rm el}(\bar{T},\pi_{\rm ca})$ on the FC curve''.
Here, $\pi_{\rm an}$ is kept at zero, and $I_{\rm el}(\bar{T},\pi_{\rm ca})$ can be explicitly obtained from \eqref{eq:P_P2G_ref}-\eqref{eq:steady_T_in}.

For the subsequent numerical case of Section IV, the MPP location process is depicted in the left-hand plot of Fig. \ref{fig:MPP_curve}.
Then, the MPP curve can be obtained from the locus of MPPs at various loading conditions (value of $P_{\rm P2G,ref}$), such as MPP$_{\rm 1}$ at $\rm 300 kW$, MPP$_{\rm 2}$ at $\rm 600 kW$, MPP$_{\rm 3}$ at $\rm 800 kW$, and so on, as shown in the right-hand plot of Fig. \ref{fig:MPP_curve}.
For a given HT-P2G system, the numerical MPP curve can be calculated in advance and stored in the MPP solver so that a look-up table can be provided during practical real-time control to reduce the computational cost.
The computed optimal parameters $(I_{\rm el,opt},\bar{T}_{\rm opt},\bm{\pi}_{\rm opt})$ play important roles in the whole MPPT control strategy, as shown in Fig. \ref{fig:MPPTControllerDiagram}.

\subsection{MPPT Controller}
\begin{figure}[!t]
  \centering
  \includegraphics[width=0.49\textwidth]{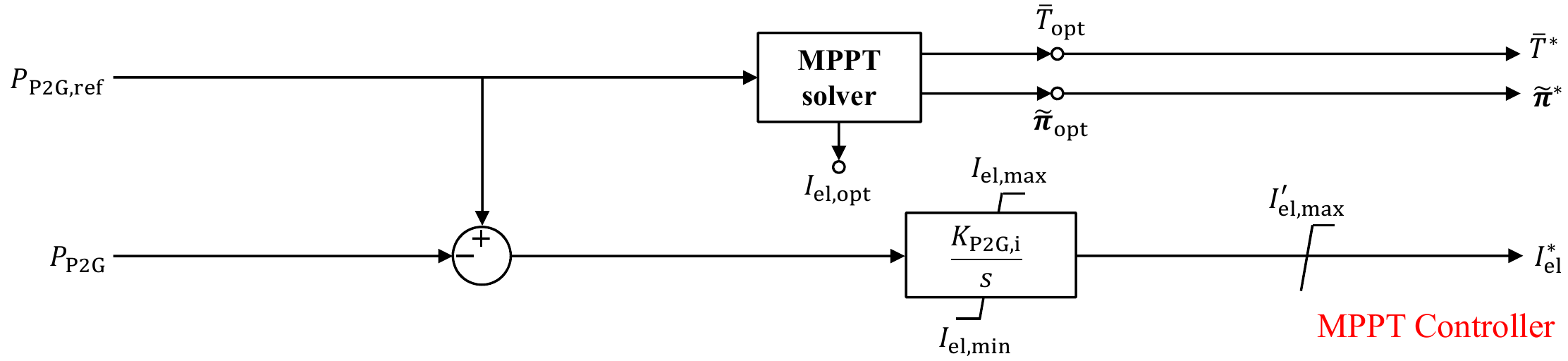}\\
  \caption{Design of the MPPT controllers.}\label{fig:MPPTControllerDiagram}
\end{figure}
As shown in Fig. \ref{fig:SystemDiagram}, the MPPT controller computes the operating parameter commands $\bar{T}^*$, $\bm{\pi}^*$ and $I^*_{\rm el}$ from the actual and reference loading power $P_{\rm P2G}$ and $P_{\rm P2G,ref}$.
The proposed design of the MPPT controller to ensure both a rapid tracking response and an improved steady-state efficiency is illustrated in Fig. \ref{fig:MPPTControllerDiagram}.
The aforementioned MPP solver is implemented here to calculate the optimal parameters $(I_{\rm el,opt},\bar{T}_{\rm opt},\bm{\pi}_{\rm opt})$, while only $\bar{T}_{\rm opt}$ and $\bm{\pi}_{\rm opt}$ are employed as $\bar{T}^*$ and $\bm{\pi}^*$, respectively.
On the other hand, the current command $I_{\rm el}^*$ is calculated by integrating the tracking error between $P_{\rm P2G}$ and $P_{\rm P2G,ref}$.
The upper limit $I'_{\rm el,max}$ in \eqref{eq:Ip_el_max} is activated when $P_{\rm fur}$ stabilizes at a nonzero value.

In this way, the fast dynamics of power converter can be used to eliminate---in a timely manner---the tracking error of the loading power, when the furnace and pumps are gradually responding to their altered set points.
In other words, the MPPT controller takes advantage of the fast current control to complement the slow responses of the temperature and flow rates, ensuring the system's dynamic performance.
In addition, an optimized steady-state system performance is achieved as a results of the updated set points from MPP solver.
Note that the electrolysis current $I_{\rm el}$ ultimately automatically stabilizes at $I_{\rm el,opt}$ because the system steady state is uniquely determined by $\bar{T}$, $\bm{\pi}$ and $P_{\rm P2G,ref}$ according to \eqref{eq:P_P2G_ref}-\eqref{eq:Ip_el_max}.

The computed $\bar{T}^*$ is implemented by the internal PI controller of furnace as shown in Fig. \ref{fig:AuxiliaryModuleDiagram}, but $I^*_{\rm el}$ and $\bm{\pi}^*$ need to be further processed in the following electrical controller and preprocess controller, as shown in Fig. \ref{fig:ControllerDiagram}.

\subsection{Electrical Controller}
\begin{figure}[!t]
	\centering
	\includegraphics[width=0.49\textwidth]{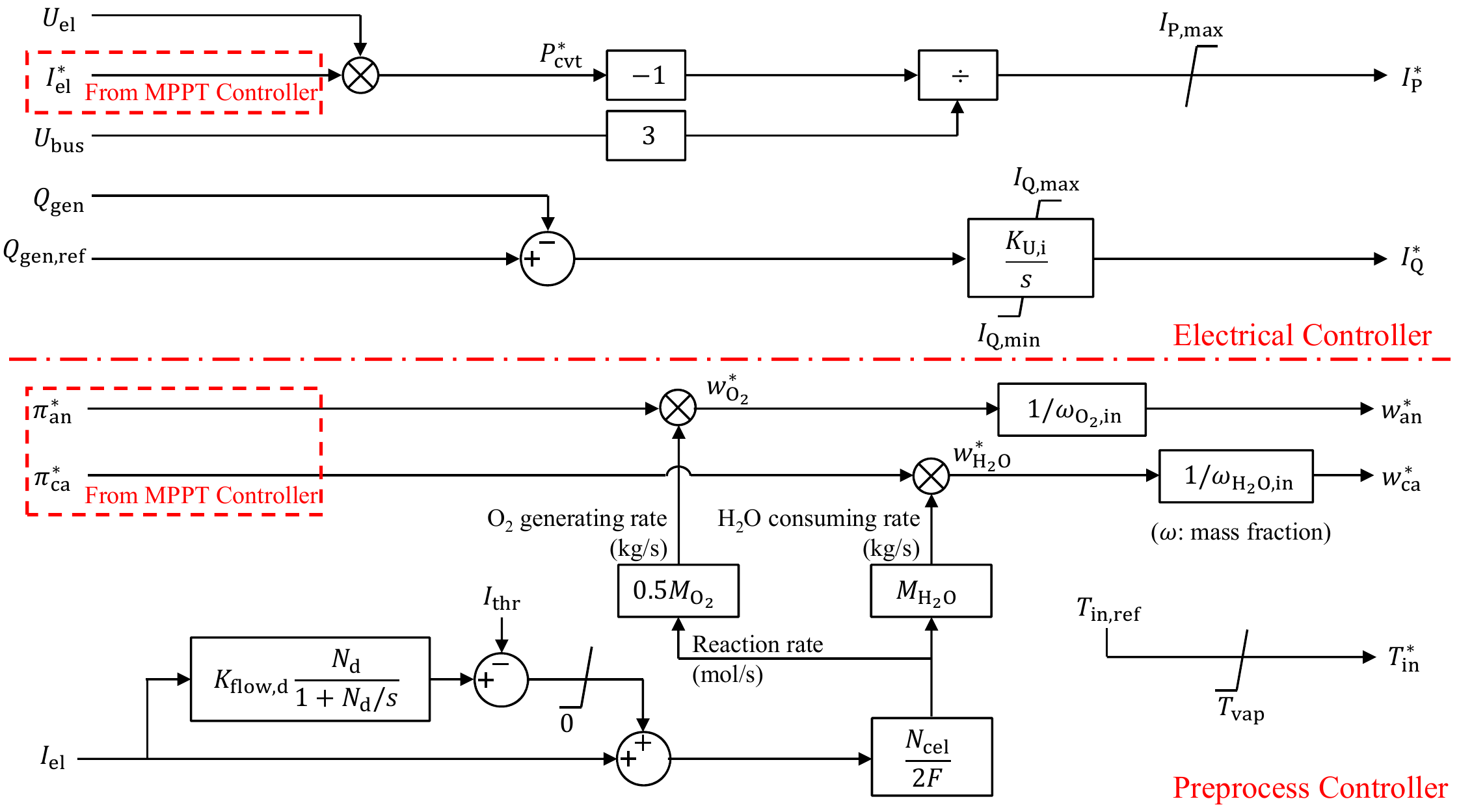}\\
	\caption{Design of the electrical controller and the preprocess controller.}\label{fig:ControllerDiagram}
\end{figure}
The commands of the active and reactive grid-side current for the power converter, $I_{\rm P}^*$ and $I_{\rm Q}^*$, are computed by the electrical controller, whose design is presented in Fig. \ref{fig:ControllerDiagram}.
This design is modified from the electrical control module of a full converter wind turbine generator \cite{Clark_GE_WindTurbine_2010}.
To provide the appropriate active power requested for electrolysis, $I_{\rm P}^*$ is calculated from the electrolysis current order $I_{\rm el}^*$ provided by the aforementioned MPPT controller based on the AC-DC active power balance.
%The monitored voltages of dc and ac sides, $U_{\rm el}$ and $U_{\rm bus}$, are also required.
On the other hand, the reactive power reference $Q_{\rm gen,ref}$ is compared to the actual reactive power generation $Q_{\rm gen}$, and the resulting error is integrated to generate $I_{\rm Q}^*$.
%Here $Q_{\rm gen,ref}$ can be set as a constant directly, or computed by specific reactive power control methods such as bus voltage control and power factor control \cite{Clark_GE_WindTurbine_2010}, which will not be elaborated here due to limited pages.

The computed $I_{\rm P}^*$ and $I_{\rm Q}^*$ are implemented by the aforementioned $dq$-current control loops and phase-locked loop of the power converter, which are represented as first-order delays of $\tau_{\rm cvt}$ in Fig. \ref{fig:AuxiliaryModuleDiagram} of Section \ref{subsec:model_auxiliary}.

\subsection{Preprocess Controller}
As shown in Fig. \ref{fig:SystemDiagram} and Fig. \ref{fig:ControllerDiagram}, the preprocess controller monitors the commands of feed factors $\pi_{\rm ca}^*$ and $\pi_{\rm an}^*$ given by the MPPT controller and restores the commands of the feed flow rates $w_{\rm ca}^*$ and $w_{\rm an}^*$ from them.
The actual electrolysis current $I_{\rm el}$ is required for computing the reaction rate used in the restoration.
The computed $w_{\rm ca}^*$ and $w_{\rm an}^*$ are implemented by the internal PI controllers of metering pumps of Section \ref{subsec:model_auxiliary}, as shown in Fig. \ref{fig:AuxiliaryModuleDiagram}.

However, if we employ the current value of $I_{\rm el}$ as the multiplier directly, feedstock starvation that damages the stack will probably occur when $I_{\rm el}$ meets a sharp rise.
Although $w_{\rm ca}^*$ and $w_{\rm an}^*$ follow the change, the actual flow rates $w_{\rm ca}$ and $w_{\rm an}$ cannot rise immediately due to the inertia of the pumps, resulting in an undesired drop of the actual feed factors $\pi_{\rm ca}$ and $\pi_{\rm an}$ and thus starvation.
Therefore, a differential block ($K_{\rm flow,d} s$) with derivative filter ($\frac{N_{\rm d}}{s+N_{\rm d}}$ to reduce the controller's sensitivity to noise \cite{Isaksson_PID_derivateFilter_2002}) is employed in Fig. \ref{fig:ControllerDiagram} to compute a correction of $I_{\rm el}$ based on the changing rate.
If the correction is larger than a specific threshold value $I_{\rm thr}$ (signifying a sharp rise in $I_{\rm el}$), this correction is incorporated into the current value of $I_{\rm el}$ to compute $I_{\rm el}'$, which can be regarded as the predicted value of $I_{\rm el}$ for $K_{\rm flow,d}$ seconds later.
This design of the preprocess controller ensures the early action of pumps to provide sufficient and timely feed flow rates $w_{\rm ca}$ and $w_{\rm an}$.

Additionally, the preprocess controller provides the temperature command for the preheater, $T_{\rm in}^*$, which is typically set as a constant higher than $T_{\rm vap}$ to ensure the vaporization of water.

\section{Case Study}
In this section, an intact HT-P2G system with a 1000-cell SOC stack is numerically modeled in the manner of Fig. \ref{fig:SystemDiagram} to validate the effects of the proposed MPPT strategy.
First, the transient process between steady states is analyzed in Section \ref{subsec:Response} as Scenarios A and B to observe the short-term response performance in tracking the load command.
Then, the system operation under a fluctuating power input is investigated in Section \ref{subsec:Fluctuating} as Scenario C to reveal the long-term production benefits.

\subsection{Case Settings}
The modeled HT-P2G system is based on the cell parameters in \cite{Udagawa_SOECmodel_2007}, the dynamic parameters in \cite{Padulles_SOFC_TransferFunction_2000}, \cite{Correa_overvoltageDynamic_2004} and \cite{Clark_GE_WindTurbine_2010}, and the necessary physical properties of involved gases.
The primary model parameters are listed in Table \ref{tab:Parameters}, and the inlet compositions are $\rm H_2O(90vol.\%)+H_2(10vol.\%)$ at the cathode and $\rm O_2(21vol.\%)+N_2(79vol.\%)$ at the anode.
Matlab-Simulink is employed to implement the model and run dynamic simulations.
%Editor: Please include both the manufacturer's name and location (including city, state, and country) for specialized equipment, software, and reagents.

\begin{table}[!t]
	\renewcommand{\arraystretch}{1.3}
	\caption{Primary model parameters of the numerical HT-P2G system}\label{tab:Parameters}
	\centering
	\begin{threeparttable}
		\begin{tabular}{lllllll}\toprule
			Parameters &Values &Units & &Parameters &Values &Units\\ \midrule
			$I_{\rm el,min}$ &53.3 &$\rm A$ & &$I_{\rm el,max}$ &480.0 &$\rm A$\\ %\hline
			$T_{\rm min}$ &750 &$\rm ^{\circ}C$ & &$T_{\rm max}$ &850 &$\rm ^{\circ}C$\\
			%$p_{\rm out}$ &1 &$\rm bar$ &$p_{\rm com}$ &100 &$\rm bar$\\
			%$T_{\rm com}$ &75 &$\rm ^{\circ}C$ &$\eta_{\rm com}$ &70\% &1 \\
			$k_{\rm rec}$ &20\% &1 & &$k_{\rm T}$ &1.085 &1 \\
			$\tau_{\rm H_2}$ &2.61 &$\rm s$ & &$\tau_{\rm H_2O}$ &3.92 &$\rm s$ \\
			$\tau_{\rm O_2}$ &2.91 &$\rm s$ & &$\tau_{\rm act}$ &0.23 &$\rm s$ \\
			$\tau_{\rm con}$ &0.23 &$\rm s$ & &$\tau_{\rm cvt}$ &0.02 &$\rm s$ \\ \bottomrule
		\end{tabular}
	\end{threeparttable}
\end{table}

The example model is simulated as a single plant connected to an infinite bus to study its dynamic response when a variation of the load command $P_{\rm P2G,ref}$ is imposed.
The bus voltage $U_{\rm bus}$ and the plant's power factor are kept at the rated values (12.66$\rm kV$) and 0.9, respectively, and $P_{\rm P2G,ref}$ is the only varying external variable.

The MPPT operation calculates the optimal temperature and feed flow rates to the control furnace and pumps that maximize the conversion efficiency according to the temporal load command $P_{\rm P2G,ref}$, as depicted by \eqref{eq:P_P2G_ref}-\eqref{eq:efficiency} and the MPP curve in Fig.\ref{fig:MPP_curve}.
For comparison, a reference operation strategy is employed as a benchmark, which sets the temperature and feed flow rates at fixed values ($\bar{T}^*\equiv 800{\rm ^{\circ}C}$, $\pi_{\rm ca}^*\equiv1.3$ and $\pi_{\rm an}^*\equiv0$ in the MPPT controller of Fig. \ref{fig:MPPTControllerDiagram}) and only controls the electrolysis current to meet the desired $P_{\rm P2G,ref}$, as depicted by the cross marker in Fig. \ref{fig:MPP_curve}.

\subsection{Transient Process between Steady States}\label{subsec:Response}
We simulated two transition scenarios for the modeled HT-P2G system.
In Scenario A, the load command $P_{\rm P2G,ref}$ of the modeled HT-P2G system steps from $400 {\rm kW}$ to $600 {\rm kW}$ at $t=100 {\rm s}$, which actually represents the transient process from MPP$_{\rm 1}$ to MPP$_{\rm 2}$ in the steady-state plot of Fig. \ref{fig:MPP_curve}.
In Scenario B, $P_{\rm P2G,ref}$ of the modeled HT-P2G system steps from $400 {\rm kW}$ to $800 {\rm kW}$ at $t=100 {\rm s}$.
The resultant transient responses of the primary variables in the subsequent $500{\rm s}$ actually represents the transient process from MPP$_{\rm 1}$ to MPP$_{\rm 3}$ in Fig. \ref{fig:MPP_curve}.
The resultant transient responses of primary variables in the subsequent $500{\rm s}$ of Scenario A and B are illustrated in Fig. \ref{fig:response_400_600} and Fig. \ref{fig:response_400_800}, respectively. 
The first 5-second process immediately after the step of Scenario A is presented in magnified form in Fig. \ref{fig:response_400_600_short}.
Based on these figures, the following effects of the proposed MPPT strategy can be observed:

\subsubsection{Rapid Tracking with Improved Efficiency (Scenario A)}
\begin{figure}[!t]
	\centering
	\includegraphics[width=0.49\textwidth]{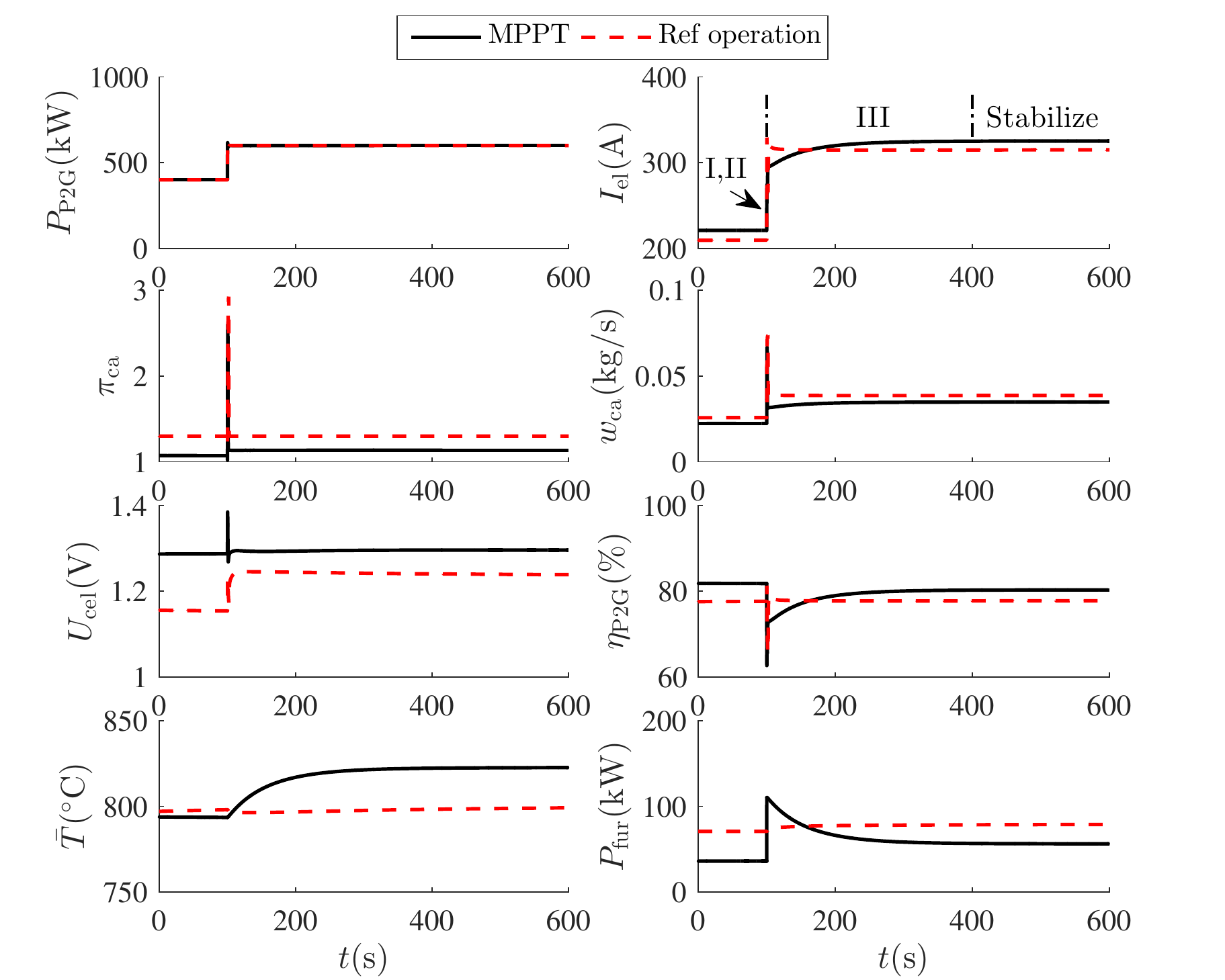}\\
	\caption{The 500-second transient response of the HT-P2G system when $P_{\rm P2G,ref}$ steps from 400$\rm kW$ to 600$\rm kW$ (MPP$_{\rm 1}$$\rightarrow$MPP$_{\rm 2}$): Scenario A.}\label{fig:response_400_600}
\end{figure}
\begin{figure}[!t]
	\centering
	\includegraphics[width=0.49\textwidth]{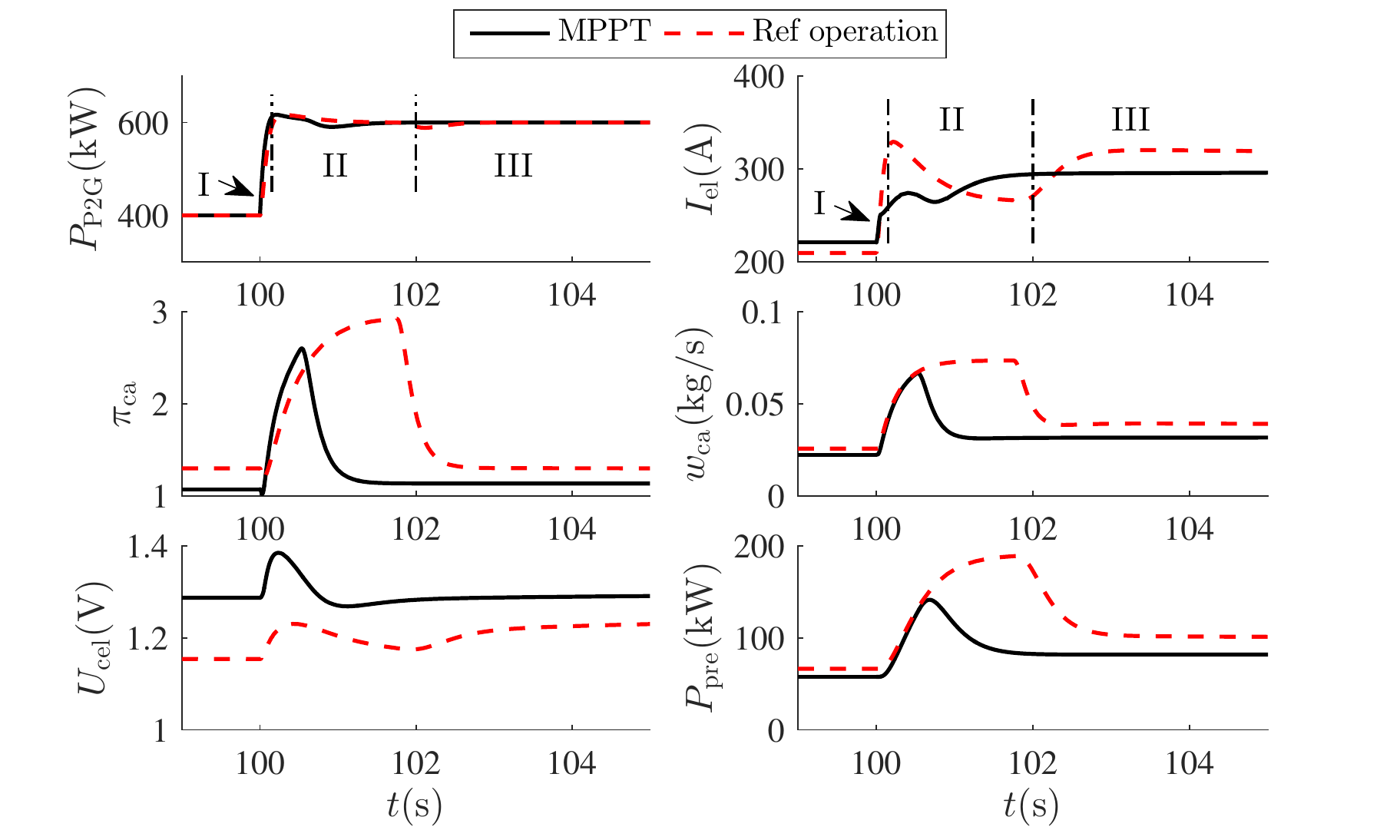}\\
	\caption{The 5-second transient response of the HT-P2G system when $P_{\rm P2G,ref}$ steps from 400$\rm kW$ to 600$\rm kW$ (MPP$_{\rm 1}$$\rightarrow$MPP$_{\rm 2}$): Magnified Scenario A.}\label{fig:response_400_600_short}
\end{figure}
As marked in Fig. \ref{fig:response_400_600} and Fig. \ref{fig:response_400_600_short}, the entire transient process of MPPT operation includes three stages before the system stabilizes at the new steady state MPP$_{\rm 2}$.
In Stage I, the imposed error between $P_{\rm P2G,ref}$ and $P_{\rm P2G}$ causes $I_{\rm el}$ to ramp up sharply as a result of fast current control of the PWM converter.
After approximately $0.15{\rm s}$, the actual load $P_{\rm P2G}$ has essentially shifted to the updated $P_{\rm P2G,ref}$, moving into Stage II.
In Stage II, the cathode feeding rate $w_{\rm ca}$ is boosted by the metering pump following the prediction control of the preprocess controller for a sharp-current-rise situation, causing simultaneous variation of $P_{\rm pre}$ for preheating and vaporization.
As a result of the rapidly controlled $I_{\rm el}$, the variation of $P_{\rm pre}$ is well compensated, and only small ripples are reflected on $P_{\rm P2G}$.
After approximately $2{\rm s}$, the effects of the boosted $w_{\rm ca}$ pass, and $I_{\rm el}$ achieves a relative stabilization but is still gradually increasing because $\bar{T}$ and $P_{\rm fur}$ are not yet stabilized. 
In Stage III, $\bar{T}$ slowly shifts towards the optimal temperature at MPP$_{\rm 2}$, and the decreasing error between $\bar{T}$ and $\bar{T}^*$ causes $P_{\rm fur}$ to decrease simultaneously.
$I_{\rm el}$ is controlled accordingly to compensate the decreasing $P_{\rm fur}$ in a much faster speed; thus, ripples of $P_{\rm P2G}$ are generally not found in this stage.
After approximately $400{\rm s}$, the whole system is essentially stabilized and ultimately arrives at the new state of MPP$_{\rm 2}$.
The features of these three stages are summarized in Table \ref{tab:Stages}.
\begin{table}[!t]
	\renewcommand{\arraystretch}{1.3}
	\caption{Three stages in the step-up system response of MPPT operation}\label{tab:Stages}
	\centering
	\begin{threeparttable}
		\begin{tabular}{llll}\toprule
			\!\!Stage &Stage I &Stage II &Stage III\\ \midrule
			\!\!Process &$P_{\rm P2G}$ responding\!\!\!\! &$P_{\rm P2G}$ stabilizing\!\!\!\! &$P_{\rm P2G}$ stabilizing\!\!\!\!\\ %\hline
			\!\!Disturbance &Stepped $P_{\rm P2G,ref}$ \!\!&Varying $P_{\rm pre}$ &Varying $P_{\rm fur}$\\
			\!\!Ripple of $P_{\rm P2G}$ \!\!\!\!\!\!&Abrupt &Small &Almost gone\\
			\!\!Time scale &Deciseconds &Seconds &Minutes\\
			\!\!Dynamic source \!\!\!\!\!\!&$I_{\rm el}$ (converter) &$w_{\rm ca}$ (pump) &$\bar{T}$ (furnace) \\ \bottomrule
		\end{tabular}
	\end{threeparttable}
\end{table}

In contrast, the transient process of reference operation in Fig. \ref{fig:response_400_600} and Fig. \ref{fig:response_400_600_short} contains only two stages: Stage I and Stage II.
In fact, the division of the three stages of the MPPT transient process is due to the naturally different time constants of $I_{\rm el}$, $w_{\rm ca}$ and $\bar{T}$.
The reference operation holds a constant temperature as shown in Fig. \ref{fig:MPP_curve}; thus, it is not surprising that Stage III is omitted.

For both operations, the actual load $P_{\rm P2G}$ shifts to the updated $P_{\rm P2G,ref}$ in approximately $0.15{\rm s}$-$0.2{\rm s}$ (Stage I), followed by small ripples lasting for approximately $2{\rm s}$-$3{\rm s}$ (Stage II), which indicates satisfactory external performance in terms of dynamic response for grid-side regulation.
As for the internal dynamic performance, the MPPT strategy has a much longer transient process than the reference operation due to the presence of Stage III.
However, as shown in Fig. \ref{fig:response_400_600}, the conversion efficiency $\eta_{\rm P2G}$ under MPPT is always higher than the reference operation at steady state.
In other words, more hydrogen is produced by the HT-P2G plant (the production rate is proportional to $I_{\rm el}$) during long-term MPPT operation.
Note that the short-lived valley of $\eta_{\rm P2G}$ is due to the temporal demand of power accumulation while shifting towards an elevated $\bar{T}^*$.

\subsubsection{Increased Loading Capacity (Scenario B)}
\begin{figure}[!t]
	\centering
	\includegraphics[width=0.49\textwidth]{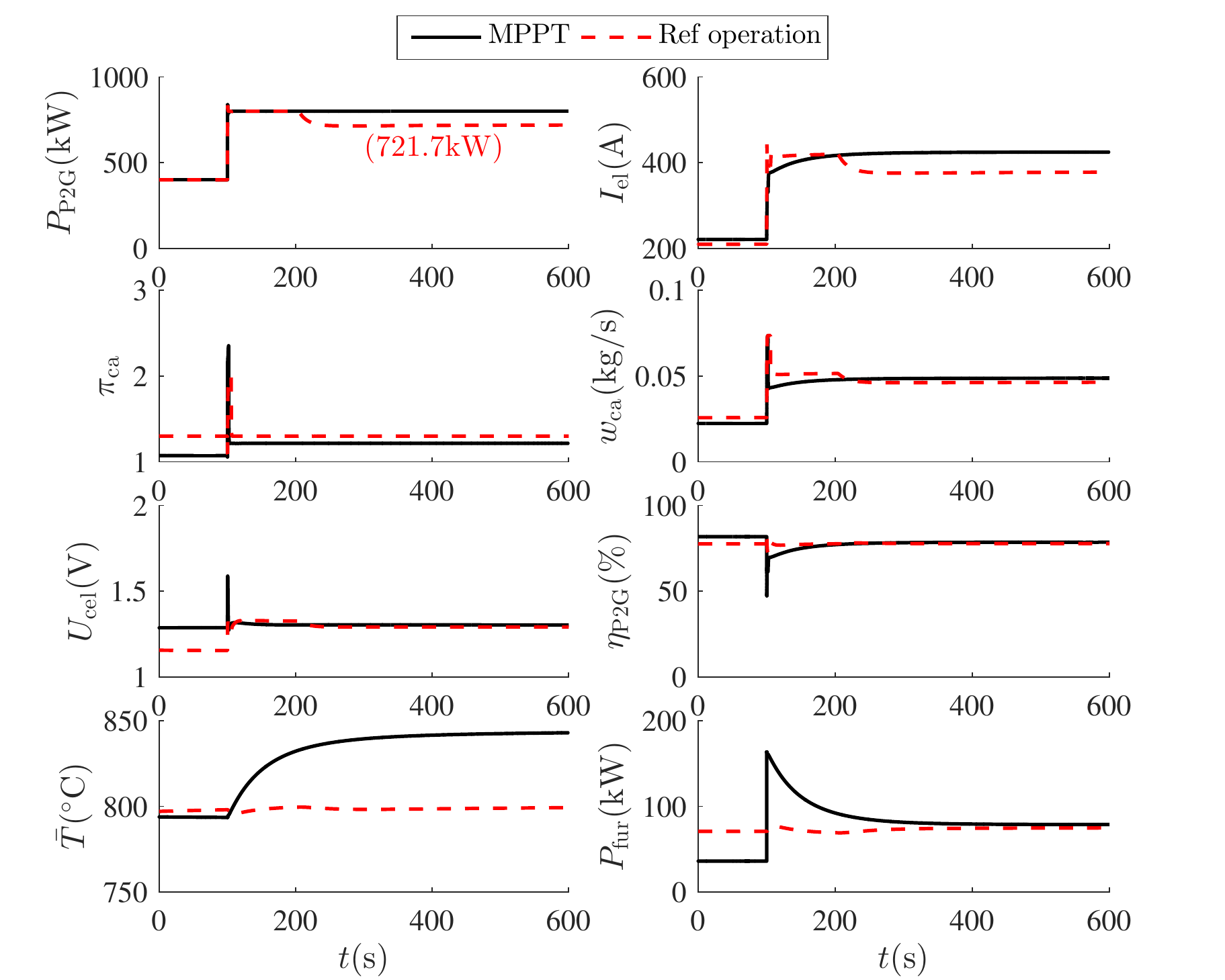}\\
	\caption{The 500-second transient response of the HT-P2G system when $P_{\rm P2G,ref}$ steps from 400$\rm kW$ to 800$\rm kW$ (MPP$_{\rm 1}$$\rightarrow$MPP$_{\rm 3}$): Scenario B.}\label{fig:response_400_800}
\end{figure}

As shown in Fig. \ref{fig:response_400_800}, the short-term dynamic performances of both MPPT and reference operations are quite similar to the step-up responses of Scenario A in Fig. \ref{fig:response_400_600} and Fig. \ref{fig:response_400_600_short}.
Nevertheless, after a short-lived tracking of $P_{\rm P2G,ref}$ at $800{\rm kW}$, the actual load $P_{\rm P2G}$ under reference operation is restricted to a maximum load of $721.7{\rm kW}$ at approximately $t=200{\rm s}$.
The curve of $I_{\rm el}$ indicates that the upper limit $I'_{\rm el,max}$ designed in the current control loop in Fig. \ref{fig:MPPTControllerDiagram}  plays an active role in this process:
the stabilized $P_{\rm fur}$ is not sufficient to ensure a stack-entrance temperature of $T_{\rm min}$ at the previous $I_{\rm el}$; thus, the current has to be restricted to $I'_{\rm el,max}$, operating at the maximum production rate affordable by $P_{\rm fur}$.
This condition does not arise in the MPPT operation at state MPP$_{\rm 3}$ because a higher $\bar{T}^*$ and a smaller $\pi_{\rm ca}^*$ have been selected to increase $I'_{\rm el,max}$ calculated by \eqref{eq:Ip_el_max}.

In fact, a closer look at the steady-state model depicted in Fig. \ref{fig:MPP_curve} indicates that the reference operation point is no longer within the feasible region (right-hand side of the $800{\rm kW}$ FC) in Scenario B.
Apart from improving efficiency, the other advantageous effect of the MPPT operation is revealed:
the steady-state loading capacity of HT-P2G can be enhanced because of the utilization of parameter feasibilities.

\subsection{Operation under Fluctuating Load Command}\label{subsec:Fluctuating}
To characterize the implementation effects of the proposed MPPT strategy under fluctuating load command, we assume the occurrence of Scenario C, where the HT-P2G plant participates in automatic generation control (AGC) as a flexible load.
Specifically, in this case, the plant receives an AGC setpoint for $P_{\rm P2G,ref}$ every 4 seconds from the grid regulator based on modified PJM AGC data \cite{Chakraborty_AGC_signal_2018}.

\begin{figure}[!t]
	\centering
	\includegraphics[width=0.49\textwidth]{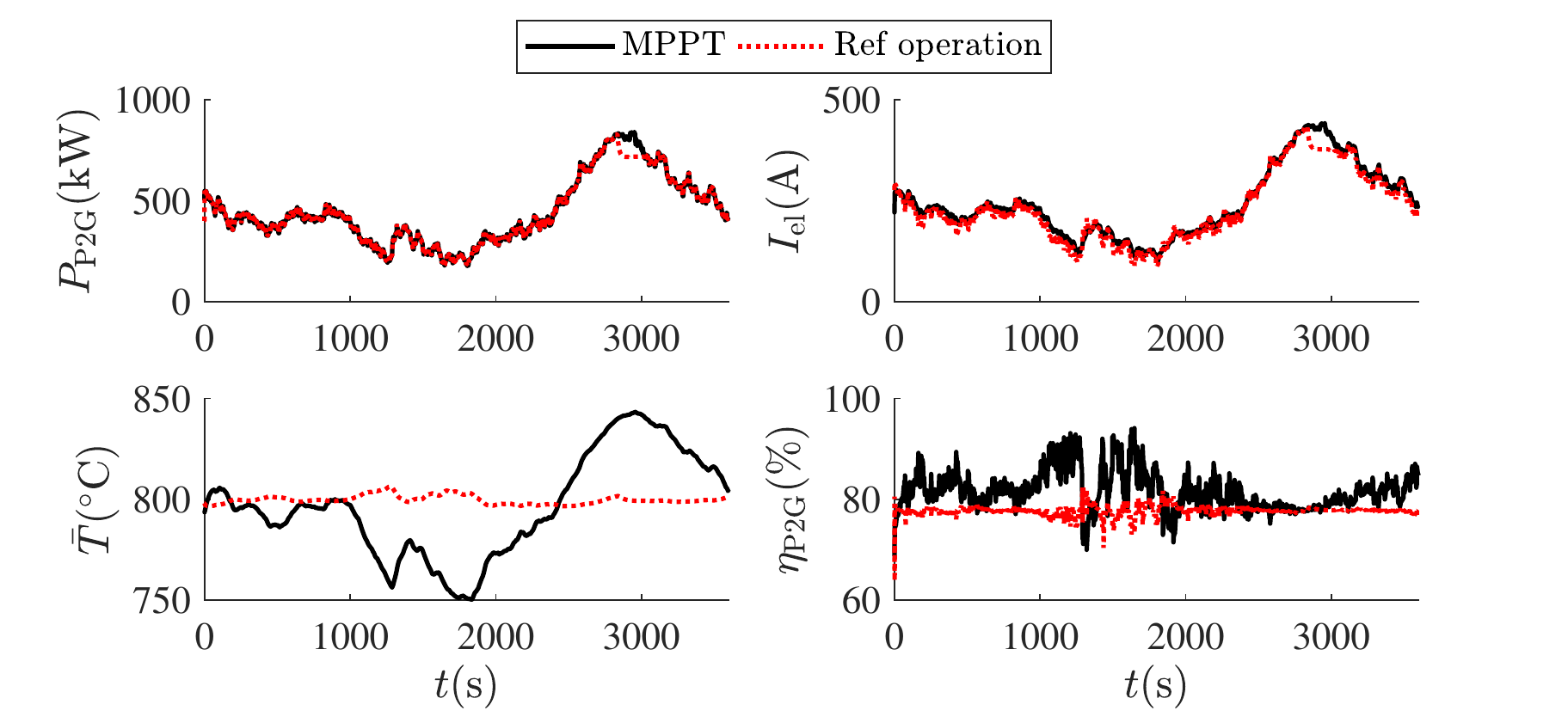}\\
	\caption{The 1-hour operation of the HT-P2G system under a fluctuating AGC signal $P_{\rm P2G,ref}$ (updated every 4 seconds): Scenario C.}\label{fig:AGC_3600}
\end{figure}

Based on the same profile of AGC signal $P_{\rm P2G,ref}$, the plant operations under both MPPT and reference strategies are simulated for 1 hour as shown in Fig. \ref{fig:AGC_3600}.
Rapid tracking of $P_{\rm P2G}$ to $P_{\rm P2G,ref}$ can be observed under both strategies except for the several minutes at approximately $t=2800{\rm s}$ when $P_{\rm P2G,ref}$ exceeds the load capacity of reference operation, and $P_{\rm P2G}$ is thus restricted to $721.7{\rm kW}$ as in Scenario B.
Moreover, the hydrogen production rate $I_{\rm el}$ and the conversion efficiency $\eta_{\rm P2G}$ under MPPT are higher than reference operation most of the time except for some temperature-ramping occasions as analyzed in Scenario A.
In brief, the MPPT effects given in Section \ref{subsec:Response} still apply.

As a natural result, a better productive performance can be achieved by MPPT relative to the reference operation under a common AGC profile, as elaborated in Table \ref{tab:Performance_AGC} with regard to this 1-hour case.
\begin{table}[!t]
	\renewcommand{\arraystretch}{1.3}
	\caption{Productive performance of the 1-hour operation under AGC}\label{tab:Performance_AGC}
	\centering
	\begin{threeparttable}
		\begin{tabular}{lll}\toprule
			Operation strategy &Reference operation &MPPT operation \\ \midrule
			Hydrogen yield &8.656$\rm kg$ &9.084$\rm kg$\\ %\hline
			Average efficiency &77.78\% &80.87\% \\ \bottomrule
		\end{tabular}
	\end{threeparttable}
\end{table}

\section{Conclusions}
By optimizing temperature and feed flow parameters of a high-temperature power-to-gas (HT-P2G) system, a maximum production point tracking (MPPT) strategy is proposed to improve the hydrogen production under variable load command given by grid regulation and then tested on a comprehensive dynamic system model.

The numerical results suggest that the short-term transient process under MPPT operation contains three stages with increasing time scales, but the load disturbances due to varying mass flow or temperature in the latter two stages are well compensated by the fast current control, and the system is essentially stabilized externally by the end of Stage I.
As a result, the MPPT-controlled plant shows a satisfying response speed for grid regulation even though the internal transition costs a few minutes.
Moreover, the steady-state conversion efficiency and available loading capacity are both increased by the MPPT strategy.
These long-term advantageous effects significantly improve the hydrogen production under specific hourly or daily loading profiles.

\appendices
\section{Derivations of Underlined Variables}\label{appsec:Underline}
As mentioned in Section \ref{subsec:MPP_solver}, the underlined variables in \eqref{eq:P_P2G_ref}-\eqref{eq:pis} can be formulated as functions of $(I_{\rm el},\bar{T},\bm{\pi})$.
In this section, the elaborated derivations are presented.

According to Fig. \ref{fig:SOCstackDiagram}, the steady-state electrolysis voltage $\underline{U_{\rm el}}$ is the sum of the reversible voltage, concentration overvoltage, activation overvoltage and ohmic overvoltage:
\setlength{\arraycolsep}{0.0em}
\begin{eqnarray}\label{eq:app_U_el}
\underline{U_{\rm el}}\left(I_{\rm el},\bar{T},\bm{\pi}\right)&{}={}&N_{\rm cel} \left[ U_{\rm rev}\left(\bar{T},\bar{\bm{p}}(\bm{\pi})\right) + U_{\rm con}\left(i_{\rm el},\bar{T},\bar{\bm{p}}(\bm{\pi})\right) \right. \nonumber\\
&&\left.+ U_{\rm act}\left(i_{\rm el},\bar{T}\right) + i_{\rm el} \cdot {\rm ASR}\left(\bar{T}\right) \right]
\end{eqnarray}
where $U_{\rm rev}(\bar{T},\bar{\bm{p}})$, $U_{\rm con}(i_{\rm el},\bar{T},\bar{\bm{p}})$, $U_{\rm act}(i_{\rm el},\bar{T})$ and $\bar{\bm{p}}(\bm{\pi})$ can be obtained from \cite{Petipas_HTE_VariousLoads_2013}.
Note that $i_{\rm el}$ is proportional to $I_{\rm el}$:
\begin{equation}\label{eq:app_i_el}
i_{\rm el} = \frac{1}{N_{\rm cha} L_x L_{y,{\rm cha}}} \cdot I_{\rm el}.
\end{equation}
\setlength{\arraycolsep}{5pt}

The steady-state power consumption of compressor $\underline{P_{\rm com}}$ can be simply derived from the compressor model in Fig. \ref{fig:AuxiliaryModuleDiagram}:
\begin{equation}\label{eq:app_P_com}
\underline{P_{\rm com}}\left(I_{\rm el}\right) = I_{\rm el} \cdot \frac{RT_{\rm com}}{2F\eta_{\rm com}} \cdot {\rm ln}\frac{p_{\rm com}}{p_{\rm out}}
\end{equation}
which is actually proportional to the compressed molar flow and the logarithm of the compression ratio.

According to the definition of feed factors $\pi_{\rm ca}$ and $\pi_{\rm an}$ as illustrated in Fig. \ref{fig:SOCstackDiagram}, the feedstock mass flows $w_{\rm ca}$ and $w_{\rm an}$ can be expressed as functions of $I_{\rm el}$ and corresponding feed factor:
\begin{equation}\label{eq:app_w_ca}
w_{\rm ca} = I_{\rm el} \cdot \frac{M_{\rm H_2O} N_{\rm cel}}{2F \omega_{\rm H_2O,in}} \cdot \pi_{\rm ca},
\end{equation}
\begin{equation}\label{eq:app_w_an}
w_{\rm an} = I_{\rm el} \cdot \frac{0.5M_{\rm O_2} N_{\rm cel}}{2F \omega_{\rm O_2,in}} \cdot \pi_{\rm an}.
\end{equation}
With the help of \eqref{eq:app_w_ca} and \eqref{eq:app_w_an}, the energy sinks contributed to feedstock warming and prewarming, $\underline{P_{\rm war}}$ and $\underline{P_{\rm war,pre}}$, can be formulated as follows:
\setlength{\arraycolsep}{0.0em}
\begin{eqnarray}\label{eq:app_P_war}
\underline{P_{\rm war}}\left(I_{\rm el},\bar{T},\bm{\pi}\right) &{}={}& \frac{I_{\rm el}N_{\rm cel}}{2F} \cdot (k_T \bar{T} - T_{\rm in}) \cdot \left(\frac{M_{\rm H_2O}}{\omega_{\rm H_2O,in}} \pi_{\rm ca} c_{\rm ca,in} \right. \nonumber \\
&&\left. + \frac{0.5M_{\rm O_2}}{\omega_{\rm O_2,in}} \pi_{\rm an} c_{\rm an,in}\right), 
\end{eqnarray}
\begin{eqnarray}\label{eq:app_P_warpre}
\underline{P_{\rm war,pre}}\left(I_{\rm el},\bm{\pi}\right) &{}={}& \frac{I_{\rm el}N_{\rm cel}}{2F} \cdot\left((T_{\rm in} - T_{\rm amb}) \frac{0.5M_{\rm O_2}}{\omega_{\rm O_2,in}} \pi_{\rm an} c_{\rm an,in} \right. \nonumber \\
&&\left.+ (T_{\rm vap} - T_{\rm amb})\frac{M_{\rm H_2O}}{\omega_{\rm H_2O,in}} \pi_{\rm ca} c_{\rm ca(liq),in} \right. \nonumber \\
&&\left.+ (T_{\rm in} - T_{\rm vap})\frac{M_{\rm H_2O}}{\omega_{\rm H_2O,in}} \pi_{\rm ca} c_{\rm ca,in}\right).
\end{eqnarray}
The equations \eqref{eq:app_P_war} and \eqref{eq:app_P_warpre} are obtained from the heat-capacity-based calculations in the warming and prewarming models of Fig. \ref{fig:EnergySinkDiagram}.
\setlength{\arraycolsep}{5pt}

As for $\underline{P_{\rm rea}}$ and $\underline{P_{\rm vap}}$, their steady-state calculations are clearly depicted in Fig. \ref{fig:EnergySinkDiagram}:
\begin{equation}\label{eq:app_P_rea}
\underline{P_{\rm rea}}\left(I_{\rm el},\bar{T}\right) = I_{\rm el} \cdot U_{\rm th}(k_T \bar{T}) \cdot N_{\rm cel},
\end{equation}
\begin{equation}\label{eq:app_P_vap}
\underline{P_{\rm vap}}\left(I_{\rm el}\right) = I_{\rm el} \cdot \frac{N_{\rm cel}}{2F} \cdot h_{\rm vap}.
\end{equation}
Note that $U_{\rm th}(k_T \bar{T})$ can be obtained from the temperature dependencies of the enthalpies of $\rm H_2O$, $\rm H_2$ and $\rm O_2$ according to the definition of the thermo-neutral voltage $U_{\rm th}$ \cite{Pan_SOECsystem_experiment_2017}.
%\section{}

%\section*{Acknowledgment}


%\ifCLASSOPTIONcaptionsoff
%  \newpage
%\fi
%
%{
%\small
%\bibliographystyle{IEEEtran}
%\bibliography{myRef}
%}

\bibliographystyle{IEEEtran}
\bibliography{IEEEabrv,myRef}

\end{document}